\documentclass[prl,twocolumn,floatfix,showpacs]{revtex4-1}
\usepackage{graphicx}
\usepackage{amsmath}
\usepackage{amssymb}
\usepackage{nccmath}

\begin{document}
\title{Probing fractional topological insulators with magnetic edge perturbations}
\author{B. B\'eri and N. R. Cooper}
\affiliation{TCM Group, Cavendish Laboratory, J.~J.~Thomson Ave., Cambridge CB3~0HE, UK}
\date{December 2011}
\begin{abstract}
We discuss detection strategies for fractional topological insulators (FTIs) realizing time-reversal invariant analogues of fractional quantum Hall systems in the Laughlin universality class. Focusing on transport measurements, we study the effect of magnetic perturbations on the edge modes. We find that the modes show unexpected robustness against magnetic backscattering for moderate couplings and edge interactions, allowing for various phase transitions signaling the FTI phase. We also describe protocols for extracting the universal integer $m$ characterizing the phase and the edge interaction parameter from the conductance of setups with magnets and a quantum point contact.
\end{abstract}
\maketitle
One of the currently most intensive fields of condensed matter research, the study of topological phases of matter\cite{HasanKane,*QiZhangRMP}, was born out of the discovery of quantum spin-Hall (QSH) insulators\cite{KaneMeleQSH1,*KaneMeleQSH2,*RoyQSH,*KonigQSH,BernevigQSHFTI}. These two dimensional gapped systems, realized in HgTe/CdTe quantum wells, display the time-reversal invariant (TRI) counterpart of the integer quantum Hall effect. Recent studies of lattice systems with topologically nontrivial flat bands\cite{Tangflatband,*Sunflatband,*Neupertflatband,*parameswaran2011fractional,*goerbig2011fractional,*murthy2011composite,*venderbos2011flat,*venderbos2011fractional}, including time-reversal invariant models\cite{NeupertFTI1,Weeks2011}, point towards realizing TRI analogues of {\it fractional} quantum Hall (FQH) systems: fractional topological insulators (FTIs)\cite{BernevigQSHFTI,LevinStern,NeupertFTI1,santos2011time}. While there are yet no predictions for concrete host  materials, the theory of FTIs has several universal aspects. One can thus already ask: if FTIs were to be realized, how one would detect them in experiments?

Our goal here is to answer this question focusing on the simplest and presumably most robust FTI phases, adiabatically connected to systems where electrons of opposite spins form opposite chirality Laughlin-like states at filling fraction $\frac{1}{m}$. The odd integer $m\!>\!1$ is the single universal parameter characterizing this phase. Note that by adiabaticity we specify only the universality class and we do not require, e.g.,  that the $z$ component of the spin is conserved by the spin-orbit coupling.

The experience with QSH and FQH systems shows that a powerful way to identify the  underlying phase is to demonstrate the existence and the universal properties of the edge modes the system supports. For QSH systems, the edge modes lead to a universally quantized two terminal conductance when the chemical potential lies in the gap\cite{HasanKane}.  In the FQHE, the zero bias (linear) tunneling conductance between the edges through a nearly pinched off quantum point contact (QPC) shows universal temperature dependence\cite{WenAW,ChangRMP} $G_\text{QPC}(T)\sim T^{2m-2}$. 
(These results are valid for temperatures, voltages, etc., much smaller than the bulk gap, which we assume throughout this paper.) 
\begin{figure}
\includegraphics[width=0.9\columnwidth,height=0.6\columnwidth]{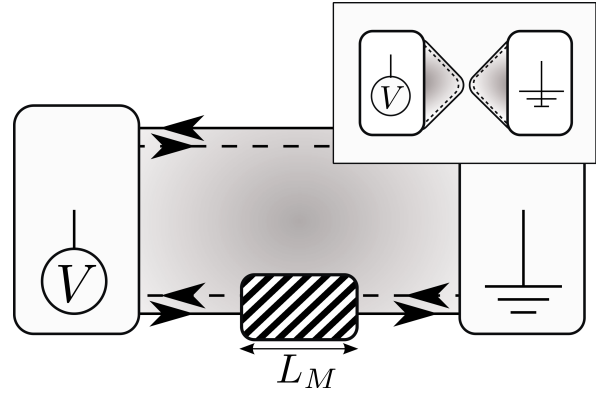}
\caption{FTI (shaded region) with a magnetic edge perturbation (hatched). The solid and dashed arrows indicate the pair of counterpropagating 
edge modes between the source (voltage $V$) and drain (grounded) terminals.  Inset: QPC setup near the pinched off limit.}
\label{fig:setups}
\end{figure}
For both the QSH and FQH cases there are fortunate circumstances which allow such universal results: 
the quantization in the QSH case holds because the contacts can be treated as Fermi liquid leads\cite{HouQPC,*TeoKaneQPC}, while 
for the $\frac{1}{m}$ FQHE, the universality of the tunneling exponent is rooted in the chirality of the edge mode\cite{KaneFQHimlett,*KaneFQHim,*Kanecontacts}. In the FTI phase the edge supports a pair of counterpropagating FQH modes and these circumstances are absent\cite{KaneFQHimlett,*KaneFQHim,*Kanecontacts}. As a result, the contact details enter the two terminal conductance, and the exponent in $G_\text{QPC}(T)$ becomes dependent on the intermode interactions. 
To extract $m$ from such a compound dependence, one has to measure a set of well chosen quantities. 
For TRI topological phases, it is natural to look for the effect of TRI {\it breaking} edge perturbations, e.g., due to magnetic fields or contacts to ferromagnetic insulators. To our knowledge, this is a direction yet unexplored in the FTI context and, as we show here, it is a fruitful one: the behavior of the conductance in the presence of such perturbations always allows one to extract $m$ from measuring at most two quantities. A sketch of the proposed setup is shown in Fig.~\ref{fig:setups}.

Our results show that the effect of magnetic perturbations is much richer than for QSH systems. While in the QSH case these  always gap the edge modes\cite{HasanKane,*QiZhangRMP}, the FTI edge is more robust:  magnetic perturbations are  {\it irrelevant} [in the sense of the renormalization group (RG)] as long as the edge interactions or the perturbation itself does not reach a critical strength.  (Here and henceforth we focus on repulsive edge interactions.) This allows for the possibility to tune the system through various phase transitions as the magnetic coupling or the edge interactions are varied, providing hallmark signatures of the FTI phase.

To begin our analysis, we summarize the relevant elements of the FTI edge theory in the absence of perturbations, following Ref.~\onlinecite{LevinStern,NeupertFTI1}. The edge can be described in terms of two bosonic quantum fields $\phi_\alpha$.
The label $\alpha=\uparrow,\downarrow$ if the $z$ component of the spin is conserved; for more general spin-orbit couplings $\alpha=L,R$ (for left/right movers). The fields obey the Kac-Moody equal-time commutation relations
\begin{equation}
[\phi_\alpha(x),\phi_\beta(y)]=(\sigma_3)_{\alpha\beta}\frac{i\pi}{m} \text{sgn}(x-y).\label{eq:kacmoody}\end{equation}
Here and henceforth $\sigma_{1,2,3}$ denote the Pauli matrices. 
The density and current of the electrons (relative to the ground state) is given by $\rho_\alpha(x)=\frac{1}{2\pi}\partial_x\phi_\alpha$ and $j_\alpha(x)=-\frac{1}{2\pi}\partial_t\phi_\alpha$, respectively. 
The Hamiltonian is
\begin{equation}
H\!=\!\int \hbar\pi mv \medop\sum_\alpha\rho_\alpha(x)\rho_\alpha(x)+\medop\sum_{\alpha\beta}V_{\alpha\beta}\rho_\alpha(x)\rho_\beta(x)dx.\end{equation}
The real positive definite matrix $V$  accounts for screened two body interactions and $v$ (with $mv\!>\!0$) is the edge velocity for $V\!=\!0$. Due to TRI, one has $V\!=\!\sigma_1V\sigma_1$\cite{NeupertFTI1}, which implies $V\!=\!u_4\openone_2\!+\!u_2\sigma_1$. 
The operator 
\begin{equation}
\Psi^\dagger_{\text{qp},\alpha}(x)\propto \exp\left[i(\sigma_3)_{\alpha\alpha}\phi_\alpha(x)\right]\label{eq:qpop}\end{equation}
creates an edge excitation of charge $\frac{1}{m}$; the electron operator is $\Psi^\dagger_{\text{e},\alpha}\propto[\Psi^\dagger_{\text{qp},\alpha}]^m$. (The precise form of Eq.~\eqref{eq:qpop} depends on Klein and regularization factors which need not be specified for our purposes.) We emphasize that Eqs. \eqref{eq:kacmoody}-\eqref{eq:qpop} represent the most general abelian FTI edge theory with a single pair of edge fields that is compatible with TRI\cite{NeupertFTI1}.

Let us now consider what happens if a magnetic perturbation is present (Fig.~\ref{fig:setups}), taken as a proximity ferromagnet of length $L_M$  for definiteness. A Zeeman-like coupling of the counterpropagating electrons reads as 
\begin{equation}
H_Z\!=\!E_Z \int_{0}^{L_M} n_3[\rho_R-\rho_L]\\+|n_\perp|[e^{i\chi}\Psi_{\text{e}R}^\dagger\Psi_{\text{e}L}+h.c.]dx,
\label{eq:HZ}
\end{equation} 
where $E_Z$ measures the strength of the perturbation\cite{kffootnote} and $\mathbf{n}$ is a unit vector related to the  magnetization. (Our analysis does not depend on the precise form of this relation.) The $n_3$ term can be dropped, as it can be eliminated by a gauge transformation that leaves the total density and current invariant. This leaves us with the $n_\perp$ term which describes backscattering.

The perturbation $\propto \Psi_{\text{e}R}^\dagger\Psi_{\text{e}L}$ is however not the most general backscattering term that can be introduced, and the presence of the ferromagnet might generate other terms as well. It is an important fact that these terms can contain only electron operators. In a $\frac{1}{m}$ FQH system, it follows from the statistical angle $\frac{\pi}{m}$ of quasiparticles that local operators can change the number of quasiparticles only by integer multiples of $m$\cite{ThoulessWu,*Einarsson}. In the systems we consider, quasiparticles come in two species related by time-reversal. They have  self-statistics angle  $\frac{\pi}{m}$ and trivial mutual statistics. This means that the number of quasiparticles can be changed only by multiples of $m$ for each species. As L/R movers belong to opposite species, quasiparticle backscattering [$\propto\Psi_{\text{qp}R}^{\dagger n}\Psi_{\text{qp}L}^n$, $n\neq 0$ mod $m$] is forbidden. In the case of the simplest spin-orbit coupling which conserves the $z$ component of the spin this reduces to the requirement that the particle number for each spin is an integer, as it should be.

The form of the $n$ electron backscattering $\Psi_{\text{e}R}^{\dagger n}\Psi_{\text{e}L}^n\propto \exp[nm(\phi_R+\phi_L)]$ ($n\in\mathbb N$) suggests the introduction of the fields $\varphi=m(\phi_R+\phi_L)$ and $\theta=\phi_R-\phi_L$. They satisfy
\begin{equation}
[\varphi(x),\theta(y)]=2\pi i \ \text{sgn}(x-y)\label{eq:llcomm}\end{equation}
and $[\varphi(x),\varphi(y)]=[\theta(x),\theta(y)]=0$, while the Hamiltonian becomes
\begin{multline}
H=\frac{\hbar}{8\pi}\int u(mK)(\partial_x\theta)^2+\frac{u}{mK}(\partial_x\varphi)^2 dx\\
+\sum_{n=1}^\infty\int_0^{L_M}a_n[ F_n e^{i (n \varphi+\chi_n)}+h.c.]dx,
\label{eq:HsLL}
\end{multline}
where $F_n$ accounts for the Klein and regularization factors. The commutator \eqref{eq:llcomm} and the quadratic part of $H$ takes the form of a spinless Luttinger liquid (sLL) with interaction parameter $mK$ and velocity $u$ with
\begin{equation}
K=\sqrt{\frac{1+\lambda_4-\lambda_2}{1+\lambda_4+\lambda_2}},\ \ u=v\sqrt{(1+\lambda_4^2)-\lambda_2^2},\end{equation}
where $\lambda_j\!=\!\frac{u_j}{\hbar\pi mv}$. The density of sLL electrons is $\rho\!=\!\frac{\partial_x\varphi}{2\pi}$. 
Note that it is $K$ that behaves as the standard Luttinger interaction parameter: $K\!<\!1$ for repulsive edge interactions ($\lambda_j\!>\!0$),  while $K\!=\!1$ corresponds to a  noninteracting edge ($\lambda_j\!=\!0$). (Even though we call  $\lambda_j=0$  ``noninteracting", electrons form a correlated FTI edge fluid even in this case.)  Eq.~\eqref{eq:HsLL} shows that any $K$ dependence will be through $mK$. 
Note also that $n$ electron backscattering processes in the original theory remain $n$ electron backscattering processes of the sLL.
Despite its simplicity, this mapping proves convenient by allowing the use of some basic results for sLLs\cite{KFluttlett,*KFluttPRB} to predict the behavior of the FTI edge - ferromagnet system. A similar mapping also appeared in the context of FQH antiwires\cite{WenAW,RennAW,KaneAW}. 

The relevance of the backscattering terms with different $n$ can be inferred from their scaling dimensions\cite{KFluttlett,*KFluttPRB} $\Delta_n=n^2mK$. The low energy properties are dominated by the process with the smallest $\Delta_n$: single electron backscattering. Our observation that quasiparticle backscattering is forbidden now becomes crucial: were it not the case, it would be quasiparticle backscattering that determines the low energy behavior. Henceforth, we focus our attention to single electron backscattering and neglect the terms with $n>1$. The remaining term is the original perturbation Eq.~\eqref{eq:HZ}, thus $a_1\propto E_Z$ and $\chi_1=\chi$.

We will analyze the effect of the ferromagnet in two opposite limits, $L_M\!\ll \!L_T$ and $L_M\!\gg\!L_T$ where $L_T\!=\!\frac{\hbar u}{k_BT}$ is the thermal length. To establish experimentally whether $L_M\!\ll\!L_T$ or $L_M\!\gg\!L_T$, one needs the value of $u$; this can be obtained for example from time-domain measurements\cite{Ashoori92}.

For $L_M\!\ll \!L_T$, the characteristic wavelength of the relevant excitations is much longer than $L_M$, hence the magnet can be taken as a delta-function impurity with strength $E_Z L_M$. The behavior of the system can be inferred following the by now classic analysis of Ref.~\onlinecite{KFluttlett,*KFluttPRB}. For a weak Zeeman term, $E_Z L_M\!\ll\! \hbar u$, the leading order RG flow of the dimensionless coupling $c_M=\frac{E_ZL_M}{\hbar u}$ is 
\begin{equation}
\frac{dc_M}{dl}=(1-mK)c_M.
\label{eq:cMRG}\end{equation}
This immediately gives our first result, that the ferromagnet is an {\it irrelevant} perturbation, as long as the repulsive interactions are moderate, $K\!>\!\frac{1}{m}$. This is in  stark contrast to the naive expectation that TRI breaking perturbation should always be relevant for a TRI topological phase. 
Eq.~\eqref{eq:cMRG} also determines the temperature dependence of the leading correction  to the linear conductance, 
\begin{equation}
\delta G(T)\propto T^{2mK-2}.\label{eq:GTDDwc}\end{equation}
The correction decays as $T\!\rightarrow\! 0$ for $K\!>\!\frac{1}{m}$, while for $K\!<\!\frac{1}{m}$ it is divergent, indicating that the flow is to strong coupling. Eq.~\eqref{eq:GTDDwc} is then valid only as long as $\frac{\delta G(T)}{G(c_M=0)}\!\ll \!1$. The behavior in the strong coupling regime corresponds to the physical picture in which the backscattering at the ferromagnet is so strong that it effectively cuts the edge into two halves. We then have two half-infinite sLLs [with fields $\varphi_j$, $\theta_j$ with $j=1(2)$ to the left (right) of the impurity], and the most relevant  process is single electron tunneling, described by\cite{KFluttlett,*KFluttPRB}
\begin{equation}
H_\text{tun}\propto d_M e^{i(\theta_1-\theta_2)/2}+h.c.\end{equation}
$H_\text{tun}$ introduces a $(-1)^{j} 2\pi$ kink in $\varphi_j$,  implementing the desired changes in the sLL charge density. Note that a $\pm 2\pi$ kink in $\varphi_j$ amounts to a $\pm \frac{2\pi}{m}$ kink in $\phi_{R j}+\phi_{L j}$, which corresponds to the transfer of charge $\frac{e}{m}$, i.e.,  quasiparticle tunneling in the physical system. 
The scaling dimension of $d_M$ is now\cite{KFluttlett,*KFluttPRB} $(mK)^{-1}$
leading to a conductance (defined without the constant contribution of the complementary edge) that decays  for $T\rightarrow 0$ as
\begin{equation}
G\propto T^{\frac{2}{mK}-2}.\label{eq:GTDDsc}
\end{equation}
   
The demonstration of the robustness of the edge modes against a ferromagnet in systems with moderate edge interactions would already provide a strong signature of the FTI phase. An even more apparent signature would be a demonstration of the phase transition upon tuning the interaction strength across $K=\frac{1}{m}$, e.g., by changing the confinement potential. 
Note that from the temperature dependence itself only the combination $mK$ can be extracted. 
In the $K<\frac{1}{m}$ case, the fact that the transport takes place through quasiparticle tunneling events can be used to obtain $m$ independently through shot noise measurements; these are however more difficult technically than measuring the conductance.

\begin{figure}
\includegraphics[width=0.9\columnwidth,height=0.4\columnwidth]{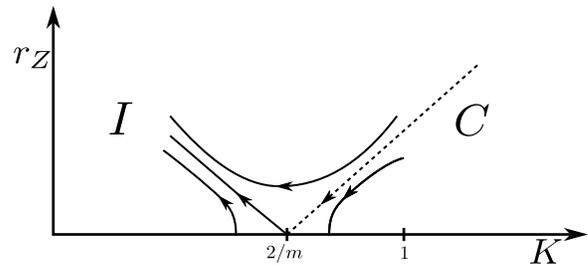}
\caption{The Kosterlitz-Thouless flow in the regime $L_M\gg L_T$ with conducting (C) and insulating (I) phases (separated by the dashed line).}
\label{fig:KTflow}
\end{figure}
Let us now turn to the opposite limit, $L_M\gg L_T$.  The system is now governed by a sine-Gordon Hamiltonian, resulting in the RG equations\cite{Giabook}
\begin{eqnarray}
\frac{dr_Z}{dl}&=&(2-mK)r_Z,\label{eq:RGEZ}\\
\frac{dK}{dl}&=&-\gamma r_Z^2K^2\label{eq:RGK},
\end{eqnarray}
to leading order in $E_Z$. Here $r_Z\!=\!\frac{E_Z}{E_c}$ is a dimensionless coupling ($E_c$ is the high energy cutoff), and $\gamma$ is a nonuniversal positive constant. 
Equations \eqref{eq:RGEZ},\eqref{eq:RGK} describe a Kosterlitz-Thouless (KT) flow (Fig.~\ref{fig:KTflow}) separating conducting and insulating phases. 
For weak Zeeman coupling and $K\!>\!\frac{2}{m}$ the magnetic perturbation is again irrelevant (conducting phase). In this regime, the leading order correction to the linear conductance obeys $\delta G\!=\!\frac{e^2}{h}\  f(r_Z (\frac{L_T}{L_c})^{2-mK},\frac{L_M}{L_T})$, with the short distance cutoff  $L_c$, where we neglected the flow of the interaction parameter $K$ as it scales only to second order in $r_Z$.
As $L_T$ defines the dephasing length\cite{KaneAW}, we expect an Ohmic behavior $\delta G\!\sim\! L_M$ for $L_M\!\gg\! L_T$. This leads to a power law decay  
\begin{equation}
\delta G(T)\propto T^{2mK-3}
\label{eq:GsG}
\end{equation}
as $T\rightarrow 0$. 
 As before, $\delta G(T)$ depends only on $mK$. 
In the complementary, strong coupling regime of Fig.~\ref{fig:KTflow} the system becomes gapped, and the conductance (of the relevant edge) is exponentially suppressed in $L_M$ (insulating phase). Reaching this regime does not require tuning $K$ below $\frac{2}{m}$; it can also be reached through the KT transition  by increasing $E_Z$. As we now show, in this regime $m$ can be  extracted independently using a setup with spatially varying  magnetization, as suggested originally in the QSH context\cite{qi2008fractional}. For large $E_Z$, the Hamiltonian can be subjected to a saddle point analysis through the imaginary time path integral for action $S=\int dx d\tau \mathcal{L}$ with
\begin{equation}
\mathcal{L}=\frac{1}{8\pi m K}[u(\partial_x\varphi)^2+u^{-1}(\partial_\tau\varphi)^2]+g_M \cos(\varphi+\chi),\end{equation}
where $g_M\propto E_Z$. For large $E_Z$ and constant $\chi$, the field $\varphi$ is locked to a  minimum of the cosine.  Now if $\chi$ varies spatially from $\chi_0$ to $\chi_{L_M}$ as one goes from $0$ to $L_M$, charge $\frac{e(\chi_{L_M}-\chi_0)}{2\pi m}$ will be accumulated along the magnetic region. To relate $\chi$ to the magnetization note that time-reversal amounts to $\chi\rightarrow\chi+\pi$. A magnetic domain wall with opposite polarizations thus corresponds to a  $0-\pi$ domain wall in $\chi$.  The trapped $\frac{e}{2m}$ charge in such a structure can be conveniently  detected through Coulomb blockade measurements\cite{qi2008fractional}.

So far we have seen that using ferromagnetic perturbations, we can measure $mK$ through the temperature dependence of the conductance. We have also shown that $m$ can be measured independently using a domain wall configuration, or noise measurements if small $K$ or large $E_Z$ can be achieved. In closing we now show that an additional conductance measurement in a QPC geometry (see Fig.~\ref{fig:setups}) provides a way to access $K$ and $m$ separately, even if small $K$ or large $E_Z$ is not reachable. We focus on the limit when the QPC is almost completely pinched off. In this case, transport is through the tunneling of electrons from the left to the right. Similarly to QSH QPCs\cite{HouQPC,*TeoKaneQPC}, this problem can now be mapped to a spinful Luttinger liquid with charge and spin interactions $K_\rho=mK$, $K_\sigma=\frac{m}{K}$. We find that for $K>\frac{1}{2m}$ charge transport is dominated by single electron tunneling processes\cite{2ppfootnote}, with scaling dimension $\Delta_e=\frac{m(K+K^{-1})}{2}$. The $T\rightarrow 0$ decay of the  linear conductance is thus
\begin{equation}
G_\text{QPC}(T)\propto T^{m(K+K^{-1})-2}.\label{eq:GQPC}\end{equation}

Our results can be used to devise protocols for measurements for all regimes of $E_Z$ and $K$. For example, starting with a $L_M\!\gg\! L_T$ setup, the conductance can either show a power law increase or an exponential suppression as $T\!\rightarrow\! 0$. The former behavior is already indicative of an FTI with $K\!>\!\frac{2}{m}$. The product  $mK$ can be measured through $\delta G(T)$, and a further measurement using the QPC setup provides $m$ and $K$ separately. In the latter case $m$ can be directly measured using a domain wall setup. The value of $K$ can also be obtained from measurements 
in the $L_M\!\ll\! L_T$ limit.

In summary, we have shown that the behavior of the edge modes under magnetic perturbations can be used to detect the FTI phase through conductance measurements. Searching for the predicted phase transitions and the temperature dependences provides a straightforward way for identifying the FTI phase in future experiments.

\acknowledgments{NRC acknowledges useful discussions with Ady Stern. This work was supported by EPSRC Grant EP/F032773/1 and BIRAX.}


\begin{thebibliography}{37}
\expandafter\ifx\csname natexlab\endcsname\relax\def\natexlab#1{#1}\fi
\expandafter\ifx\csname bibnamefont\endcsname\relax
  \def\bibnamefont#1{#1}\fi
\expandafter\ifx\csname bibfnamefont\endcsname\relax
  \def\bibfnamefont#1{#1}\fi
\expandafter\ifx\csname citenamefont\endcsname\relax
  \def\citenamefont#1{#1}\fi
\expandafter\ifx\csname url\endcsname\relax
  \def\url#1{\texttt{#1}}\fi
\expandafter\ifx\csname urlprefix\endcsname\relax\def\urlprefix{URL }\fi
\providecommand{\bibinfo}[2]{#2}
\providecommand{\eprint}[2][]{\url{#2}}

\bibitem[{\citenamefont{Hasan and Kane}(2010)}]{HasanKane}
\bibinfo{author}{\bibfnamefont{M.~Z.} \bibnamefont{Hasan}} \bibnamefont{and}
  \bibinfo{author}{\bibfnamefont{C.~L.} \bibnamefont{Kane}},
  \bibinfo{journal}{Rev. Mod. Phys.} \textbf{\bibinfo{volume}{82}},
  \bibinfo{pages}{3045} (\bibinfo{year}{2010}); 
\bibitem[{\citenamefont{Qi and Zhang}(2011)}]{QiZhangRMP}
\bibinfo{author}{\bibfnamefont{X.-L.} \bibnamefont{Qi}} \bibnamefont{and}
  \bibinfo{author}{\bibfnamefont{S.-C.} \bibnamefont{Zhang}},
  \bibinfo{journal}{Rev. Mod. Phys.} \textbf{\bibinfo{volume}{83}},
  \bibinfo{pages}{1057} (\bibinfo{year}{2011}). 
\bibitem[{\citenamefont{Kane and Mele}(2005{\natexlab{a}})}]{KaneMeleQSH1}
\bibinfo{author}{\bibfnamefont{C.~L.} \bibnamefont{Kane}} \bibnamefont{and}
  \bibinfo{author}{\bibfnamefont{E.~J.} \bibnamefont{Mele}},
  \bibinfo{journal}{Phys. Rev. Lett.} \textbf{\bibinfo{volume}{95}},
  \bibinfo{pages}{146802} (\bibinfo{year}{2005}{\natexlab{a}}); 
\bibitem[{\citenamefont{Kane and Mele}(2005{\natexlab{b}})}]{KaneMeleQSH2}
\bibinfo{author}{\bibfnamefont{C.~L.} \bibnamefont{Kane}} \bibnamefont{and}
  \bibinfo{author}{\bibfnamefont{E.~J.} \bibnamefont{Mele}},
  \bibinfo{journal}{Phys. Rev. Lett.} \textbf{\bibinfo{volume}{95}},
   \bibinfo{pages}{226801} (\bibinfo{year}{2005}{\natexlab{b}}); 
\bibitem[{\citenamefont{Roy}(2009)}]{RoyQSH}
\bibinfo{author}{\bibfnamefont{R.}~\bibnamefont{Roy}}, \bibinfo{journal}{Phys.
  Rev. B} \textbf{\bibinfo{volume}{79}}, \bibinfo{pages}{195321}
  (\bibinfo{year}{2009}); 
\bibitem[{\citenamefont{\relax{K\"onig \it et al.}}(2007)}]{KonigQSH}
\bibinfo{author}{\bibfnamefont{M.}~\bibnamefont{\relax{K\"onig \it et al.}}},
  \bibinfo{journal}{Science} \textbf{\bibinfo{volume}{318}},
   \bibinfo{pages}{766} (\bibinfo{year}{2007}). 
\bibitem[{\citenamefont{Bernevig and Zhang}(2006)}]{BernevigQSHFTI}
\bibinfo{author}{\bibfnamefont{B.~A.} \bibnamefont{Bernevig}} \bibnamefont{and}
  \bibinfo{author}{\bibfnamefont{S.-C.} \bibnamefont{Zhang}},
  \bibinfo{journal}{Phys. Rev. Lett.} \textbf{\bibinfo{volume}{96}},
  \bibinfo{pages}{106802} (\bibinfo{year}{2006}). 
\bibitem[{\citenamefont{{\relax Tang {\it et al}}}(2011)}]{Tangflatband}
\bibinfo{author}{\bibfnamefont{E.}~\bibnamefont{{\relax Tang {\it et al}}}},
  \bibinfo{journal}{Phys. Rev. Lett.} \textbf{\bibinfo{volume}{106}},
  \bibinfo{pages}{236802} (\bibinfo{year}{2011}); 
\bibitem[{\citenamefont{{\relax Sun {\it et al.}}}(2011)}]{Sunflatband}
\bibinfo{author}{\bibfnamefont{K.}~\bibnamefont{{\relax Sun {\it et al.}}}},
  \bibinfo{journal}{Phys. Rev. Lett.} \textbf{\bibinfo{volume}{106}},
  \bibinfo{pages}{236803} (\bibinfo{year}{2011}); 
\bibitem[{\citenamefont{{\relax Neupert {\it et al.}}}(2011)}]{Neupertflatband}
\bibinfo{author}{\bibfnamefont{T.}~\bibnamefont{{\relax Neupert {\it et
  al.}}}}, \bibinfo{journal}{Phys. Rev. Lett.} \textbf{\bibinfo{volume}{106}},
  \bibinfo{pages}{236804} (\bibinfo{year}{2011}); 
\bibitem[{\citenamefont{Parameswaran et~al.}(2011)\citenamefont{Parameswaran,
  Roy, and Sondhi}}]{parameswaran2011fractional}
\bibinfo{author}{\bibfnamefont{S.~A.} \bibnamefont{Parameswaran}},
  \bibinfo{author}{\bibfnamefont{R.}~\bibnamefont{Roy}}, \bibnamefont{and}
  \bibinfo{author}{\bibfnamefont{S.~L.} \bibnamefont{Sondhi}},
  \bibinfo{journal}{arXiv:1106.4025}  (\bibinfo{year}{2011}); 
\bibitem[{\citenamefont{Goerbig}(2011)}]{goerbig2011fractional}
\bibinfo{author}{\bibfnamefont{M.}~\bibnamefont{Goerbig}},
  \bibinfo{journal}{arXiv:1107.1986}  (\bibinfo{year}{2011}); 
\bibitem[{\citenamefont{Murthy and Shankar}(2011)}]{murthy2011composite}
\bibinfo{author}{\bibfnamefont{G.}~\bibnamefont{Murthy}} \bibnamefont{and}
  \bibinfo{author}{\bibfnamefont{R.}~\bibnamefont{Shankar}},
  \bibinfo{journal}{arXiv:1108.5501}  (\bibinfo{year}{2011}); 
\bibitem[{\citenamefont{Venderbos et~al.}(2011)\citenamefont{Venderbos,
  Daghofer, and v-d Brink}}]{venderbos2011flat}
\bibinfo{author}{\bibfnamefont{J.}~\bibnamefont{Venderbos}},
  \bibinfo{author}{\bibfnamefont{M.}~\bibnamefont{Daghofer}}, \bibnamefont{and}
  \bibinfo{author}{\bibfnamefont{J.}~\bibnamefont{v-d Brink}},
  \bibinfo{journal}{Phys. Rev. Lett.} \textbf{\bibinfo{volume}{107}},
  \bibinfo{pages}{116401} (\bibinfo{year}{2011}); 
\bibitem[{\citenamefont{{\relax Venderbos {\it et
  al}}}(2011)}]{venderbos2011fractional}
\bibinfo{author}{\bibfnamefont{J.}~\bibnamefont{{\relax Venderbos {\it et
  al}}}}, \bibinfo{journal}{arXiv:1109.5955}  (\bibinfo{year}{2011}). 
\bibitem[{\citenamefont{Neupert et~al.}(2011)\citenamefont{Neupert, Santos,
  Ryu, Chamon, and Mudry}}]{NeupertFTI1}
\bibinfo{author}{\bibfnamefont{T.}~\bibnamefont{Neupert}},
  \bibinfo{author}{\bibfnamefont{L.}~\bibnamefont{Santos}},
  \bibinfo{author}{\bibfnamefont{S.}~\bibnamefont{Ryu}},
  \bibinfo{author}{\bibfnamefont{C.}~\bibnamefont{Chamon}}, \bibnamefont{and}
  \bibinfo{author}{\bibfnamefont{C.}~\bibnamefont{Mudry}},
  \bibinfo{journal}{Phys. Rev. B} \textbf{\bibinfo{volume}{84}},
  \bibinfo{pages}{165107} (\bibinfo{year}{2011}). 
\bibitem[{\citenamefont{Weeks and Franz}(2011)}]{Weeks2011}
\bibinfo{author}{\bibfnamefont{C.}~\bibnamefont{Weeks}} \bibnamefont{and}
  \bibinfo{author}{\bibfnamefont{M.}~\bibnamefont{Franz}},
  \bibinfo{journal}{arXiv:1111.1447}  (\bibinfo{year}{2011}). 
\bibitem[{\citenamefont{Levin and Stern}(2009)}]{LevinStern}
\bibinfo{author}{\bibfnamefont{M.}~\bibnamefont{Levin}} \bibnamefont{and}
  \bibinfo{author}{\bibfnamefont{A.}~\bibnamefont{Stern}},
  \bibinfo{journal}{Phys. Rev. Lett.} \textbf{\bibinfo{volume}{103}},
  \bibinfo{pages}{196803} (\bibinfo{year}{2009}). 
\bibitem[{\citenamefont{Santos et~al.}(2011)\citenamefont{Santos, Neupert, Ryu,
  Chamon, and Mudry}}]{santos2011time}
\bibinfo{author}{\bibfnamefont{L.}~\bibnamefont{Santos}},
  \bibinfo{author}{\bibfnamefont{T.}~\bibnamefont{Neupert}},
  \bibinfo{author}{\bibfnamefont{S.}~\bibnamefont{Ryu}},
  \bibinfo{author}{\bibfnamefont{C.}~\bibnamefont{Chamon}}, \bibnamefont{and}
  \bibinfo{author}{\bibfnamefont{C.}~\bibnamefont{Mudry}},
  \bibinfo{journal}{Phys. Rev. B} \textbf{\bibinfo{volume}{84}},
  \bibinfo{pages}{165138} (\bibinfo{year}{2011}). 
\bibitem[{\citenamefont{Wen}(1991)}]{WenAW}
\bibinfo{author}{\bibfnamefont{X.-G.} \bibnamefont{Wen}},
  \bibinfo{journal}{Phys. Rev. B} \textbf{\bibinfo{volume}{44}},
  \bibinfo{pages}{5708} (\bibinfo{year}{1991}). 
\bibitem[{\citenamefont{Chang}(2003)}]{ChangRMP}
\bibinfo{author}{\bibfnamefont{A.~M.} \bibnamefont{Chang}},
  \bibinfo{journal}{Rev. Mod. Phys.} \textbf{\bibinfo{volume}{75}},
  \bibinfo{pages}{1449} (\bibinfo{year}{2003}). 
\bibitem[{\citenamefont{Hou et~al.}(2009)\citenamefont{Hou, Kim, and
  Chamon}}]{HouQPC}
\bibinfo{author}{\bibfnamefont{C.-Y.} \bibnamefont{Hou}},
  \bibinfo{author}{\bibfnamefont{E.-A.} \bibnamefont{Kim}}, \bibnamefont{and}
  \bibinfo{author}{\bibfnamefont{C.}~\bibnamefont{Chamon}},
  \bibinfo{journal}{Phys. Rev. Lett.} \textbf{\bibinfo{volume}{102}},
  \bibinfo{pages}{076602} (\bibinfo{year}{2009}); 
\bibitem[{\citenamefont{Teo and Kane}(2009)}]{TeoKaneQPC}
\bibinfo{author}{\bibfnamefont{J.~C.~Y.} \bibnamefont{Teo}} \bibnamefont{and}
  \bibinfo{author}{\bibfnamefont{C.~L.} \bibnamefont{Kane}},
  \bibinfo{journal}{Phys. Rev. B} \textbf{\bibinfo{volume}{79}},
  \bibinfo{pages}{235321} (\bibinfo{year}{2009}). 
\bibitem[{\citenamefont{Kane et~al.}(1994)\citenamefont{Kane, Fisher, and
  Polchinski}}]{KaneFQHimlett}
\bibinfo{author}{\bibfnamefont{C.~L.} \bibnamefont{Kane}},
  \bibinfo{author}{\bibfnamefont{M.~P.~A.} \bibnamefont{Fisher}},
  \bibnamefont{and}
  \bibinfo{author}{\bibfnamefont{J.}~\bibnamefont{Polchinski}},
  \bibinfo{journal}{Phys. Rev. Lett.} \textbf{\bibinfo{volume}{72}},
  \bibinfo{pages}{4129} (\bibinfo{year}{1994}); 
\bibitem[{\citenamefont{Kane and Fisher}(1995{\natexlab{a}})}]{KaneFQHim}
\bibinfo{author}{\bibfnamefont{C.~L.} \bibnamefont{Kane}} \bibnamefont{and}
  \bibinfo{author}{\bibfnamefont{M.~P.~A.} \bibnamefont{Fisher}},
  \bibinfo{journal}{Phys. Rev. B} \textbf{\bibinfo{volume}{51}},
  \bibinfo{pages}{13449} (\bibinfo{year}{1995}{\natexlab{a}}); 
\bibitem[{\citenamefont{Kane and Fisher}(1995{\natexlab{b}})}]{Kanecontacts}
\bibinfo{author}{\bibfnamefont{C.~L.} \bibnamefont{Kane}} \bibnamefont{and}
  \bibinfo{author}{\bibfnamefont{M.~P.~A.} \bibnamefont{Fisher}},
  \bibinfo{journal}{Phys. Rev. B} \textbf{\bibinfo{volume}{52}},
  \bibinfo{pages}{17393} (\bibinfo{year}{1995}{\natexlab{b}}). 
\bibitem[{kff()}]{kffootnote}
\bibinfo{note}{If the edge electrons have nonzero Fermi wavevector $k_F$, $E_Z$
  is the $2k_F$ component of the perturbation.} 
\bibitem[{\citenamefont{Thouless and Wu}(1985)}]{ThoulessWu}
\bibinfo{author}{\bibfnamefont{D.~J.} \bibnamefont{Thouless}} \bibnamefont{and}
  \bibinfo{author}{\bibfnamefont{Y.-S.} \bibnamefont{Wu}},
  \bibinfo{journal}{Phys. Rev. B} \textbf{\bibinfo{volume}{31}},
  \bibinfo{pages}{1191} (\bibinfo{year}{1985}); 
\bibitem[{\citenamefont{Einarsson}(1990)}]{Einarsson}
\bibinfo{author}{\bibfnamefont{T.}~\bibnamefont{Einarsson}},
  \bibinfo{journal}{Phys. Rev. Lett.} \textbf{\bibinfo{volume}{64}},
  \bibinfo{pages}{1995} (\bibinfo{year}{1990}). 
\bibitem[{\citenamefont{Kane and Fisher}(1992{\natexlab{a}})}]{KFluttlett}
\bibinfo{author}{\bibfnamefont{C.~L.} \bibnamefont{Kane}} \bibnamefont{and}
  \bibinfo{author}{\bibfnamefont{M.~P.~A.} \bibnamefont{Fisher}},
  \bibinfo{journal}{Phys. Rev. Lett.} \textbf{\bibinfo{volume}{68}},
  \bibinfo{pages}{1220} (\bibinfo{year}{1992}{\natexlab{a}}); 
\bibitem[{\citenamefont{Kane and Fisher}(1992{\natexlab{b}})}]{KFluttPRB}
\bibinfo{author}{\bibfnamefont{C.~L.} \bibnamefont{Kane}} \bibnamefont{and}
  \bibinfo{author}{\bibfnamefont{M.~P.~A.} \bibnamefont{Fisher}},
  \bibinfo{journal}{Phys. Rev. B} \textbf{\bibinfo{volume}{46}},
  \bibinfo{pages}{15233} (\bibinfo{year}{1992}{\natexlab{b}}). 
\bibitem[{\citenamefont{Renn and Arovas}(1995)}]{RennAW}
\bibinfo{author}{\bibfnamefont{S.~R.} \bibnamefont{Renn}} \bibnamefont{and}
  \bibinfo{author}{\bibfnamefont{D.~P.} \bibnamefont{Arovas}},
  \bibinfo{journal}{Phys. Rev. B} \textbf{\bibinfo{volume}{51}},
  \bibinfo{pages}{16832} (\bibinfo{year}{1995}). 
\bibitem[{\citenamefont{Kane and Fisher}(1997)}]{KaneAW}
\bibinfo{author}{\bibfnamefont{C.~L.} \bibnamefont{Kane}} \bibnamefont{and}
  \bibinfo{author}{\bibfnamefont{M.~P.~A.} \bibnamefont{Fisher}},
  \bibinfo{journal}{Phys. Rev. B} \textbf{\bibinfo{volume}{56}},
  \bibinfo{pages}{15231} (\bibinfo{year}{1997}). 
\bibitem[{\citenamefont{{\relax Ashoori, R. C. {\it et
  al.}}}(1992)}]{Ashoori92}
\bibinfo{author}{\bibnamefont{{\relax Ashoori, R. C. {\it et al.}}}},
  \bibinfo{journal}{Phys. Rev. B} \textbf{\bibinfo{volume}{45}},
  \bibinfo{pages}{3894} (\bibinfo{year}{1992}). 
\bibitem[{\citenamefont{Giamarchi}(2004)}]{Giabook}
\bibinfo{author}{\bibfnamefont{T.}~\bibnamefont{Giamarchi}},
  \emph{\bibinfo{title}{Quantum physics in one dimension}}, vol.
  \bibinfo{volume}{121} (\bibinfo{publisher}{Oxford University Press, USA},
  \bibinfo{year}{2004}). 
\bibitem[{\citenamefont{Qi et~al.}(2008)\citenamefont{Qi, Hughes, and
  Zhang}}]{qi2008fractional}
\bibinfo{author}{\bibfnamefont{X.}~\bibnamefont{Qi}},
  \bibinfo{author}{\bibfnamefont{T.}~\bibnamefont{Hughes}}, \bibnamefont{and}
  \bibinfo{author}{\bibfnamefont{S.}~\bibnamefont{Zhang}},
  \bibinfo{journal}{Nature Physics} \textbf{\bibinfo{volume}{4}},
  \bibinfo{pages}{273} (\bibinfo{year}{2008}). 
\bibitem[{2pp()}]{2ppfootnote}
\bibinfo{note}{For $K<\frac{1}{2m}$ two electron processes become more relevant
  with scaling dimension $\Delta_{2e}=\frac{m}{2K}$, leading to a
  $G_\text{QPC}(T)\propto T^{m/K-2}$ decay.} 
\end{thebibliography}
\end{document}